  \documentclass[10pt,letterpaper,conference]{IEEEtran}

\def\figs{7 cm}

\usepackage{epsf}
\usepackage{ifthen}
\usepackage{times}

\newcommand{\thd}[1]
{\ifthenelse {\equal{#1}{1}}
	{{#1}$^{\mathrm{st}}$} 
	{{\ifthenelse{\equal{#1}{2}}{{#1}$^{\mathrm{nd}}$}{{#1}$^{\mathrm{th}}$}}}}

\newcommand{\var}{{\bf var}}

\newcommand{\SNR}{\text{SNR}\,}

\newcommand{\const}{\text{const}}

\newcommand{\mmse}{\text{mmse}}
\newcommand{\cov}{\text{cov}}

\usepackage{amsmath}
\usepackage{amssymb}

\newtheorem{Theorem}{Theorem}

\begin{document}
\title{Spreading Signals in the Wideband Limit}
\author{Elchanan Zwecher and Dana Porrat\\
The Hebrew University of Jerusalem\\
Jerusalem, Israel 91904\\
elchanzw, dporrat@cs.huji.ac.il}
\maketitle
\begin{abstract}
Wideband communications are impossible with signals that are spread over a very large band and are transmitted over multipath channels unknown ahead of time.
This work exploits the I-mmse connection to bound the achievable data-rate of spreading signals in wideband settings, and to conclude that the achievable data-rate diminishes as the bandwidth increases due to channel uncertainty.
The result applies to all spreading modulations, i.e.\ signals that are evenly spread over the bandwidth available to the communication system, with SNR smaller than log(W/L)/(W/L) and holds for communications over channels where the number of paths L is unbounded by sub-linear in the bandwidth W.

\end{abstract}

\section{Introduction}

This work analyzes the performance of wideband communication systems, we characterize the SNR regime that does not allow communications in the wideband limit.
The underlying reason for the inability to communicate is channel uncertainty, and our result applies to channels where the number of multipath components is unbounded and sub-linear in the bandwidth.
Our proof is based on the I-mmse connection~\cite{guo_2005} and  hinges on a calculation of the minimum mean square error (mmse) estimate of the unknown channel.

We consider signals that spread their power over the entire bandwidth, such as PPM or impulse radio~\cite{win_1998}, where the pulse shape and duration determine the bandwidth, or direct sequence spread spectrum, where the chip duration determines the frequency spread of the signal.
Our result applies also to OFDM-type signals, if the entire available bandwidth is used concurrently.
Examples of signals that are not spread over bandwidth are FSK and multi-tone FSK~\cite{luo_Frequency_2002,luo_2003}, where each symbol concentrates power on a small span of frequencies, although the entire range of symbols may span a very large bandwidth.

Our result shows that channel uncertainty is detrimental to spreading systems operating over multipath channels where the number of apparent paths is unbounded but sub-linear with the system bandwidth~\cite{saadane_2004,saadane_2005}, essentially because of the signal uses too many eigen-modes of the channel.
Our model of channel variation in time is a simplistic block coherent one, where the channel is fixed for known lengths of time (coherence periods) and realizations over different coherence periods are IID.
This channel model offers the advantage of eigen-modes that are particularly simple, as harmonic signals are eigen-modes of any linear time invariant channel.
The channel uncertainty a communication system faces when operating over the block--coherent channel is thus limited to the eigen-values of the channel, or in other words to the complex channel gain over the frequency band the system uses.

The essential feature of spreading signals that renders them ineffective over wide bands, is that they use the entire range of channel eigen-modes concurrently, and are thus exposed to uncertainty of a large number of parameters (channel eigen-values).
Modulation schemes of the FSK type, that exploit a small number of channel eigen-modes per symbol, are exposed to uncertainty in only a small number of parameters.

Our result can be extended to more complex channels, where the variation in time is described using the Doppler spectrum rather than by block--coherence~\cite{raghavan_2007,sayeed_2004,durisi_2006}.
The eigen-modes of such channels are approximately given by orthogonal Weyl-Heisenberg bases~\cite{kozek_1997} in the under-spread case, i.e.\ when the channel's response is highly concentrated in the delay--Doppler plain.
The essential feature that determines whether communications are possible in the wideband limit is the spreading of symbol power over the unknown channel eigen-modes.
In channels where the eigen-modes are not known in advance, it is very difficult to overcome channel uncertainty.

Related work has shown that pulse-position modulation (PPM) systems are unable to handle uncertainty in the delays of multipath components \cite{porrat_2006,porrat_2007}.
We showed that the receiver is unable to detect the channel paths if the bandwidth is large enough, whether it uses a threshold detector~\cite{porrat_2006} or a maximum likelihood detector~\cite{porrat_2007}.
This work extends the scope of past results to a wider family of signals, and makes a statement on the achievable rate.

The comparison of our result to that of {\em Telatar \& Tse}~\cite{telatar_2000} is also interesting.
The {\em T\&T} results discussed continuous signals, i.e.\ effectively using a signal to noise ratio (SNR) that inversely depends on the system bandwidth $\SNR=\theta\left(\frac{1}{W}\right)$, whereas our result is more general in the sense that it specifies the range of SNR dependencies on bandwidth that does not allow communications in the wideband regime.
In other words, our result applies to impulsive systems and implicitly indicates the minimal level of impulsiveness that allows communications in the limit of large bandwidth.

Another point of comparison of our result to that of~\cite{telatar_2000} is the type of signals to which they apply.
{\em T\&T} consider two types of spreading signals: IID complex signals and signals with a very low cross-correlation: $\sum_i X_i X_{i-n}\sim\const$ that does not depend on the length of the vector.
Our result applies to a wider family of signals, where the cross correlation may as high as $\theta\left({\sqrt{W}}\right)$.
This is significant because spreading signals generated by IID or pseudo-random sequences have an empirical correlation that varies as $\theta\left(\sqrt{W}\right)$.

The results in~\cite{porrat_2007}, that connect the number of channel paths to the level of impulsiveness are also relevant here. 
The impulsiveness parameter there is easily translated to the SNR per active burst of transmission, and the results in~\cite{porrat_2007} basically determine a lower bound on the SNR, above which the channel uncertainty penalty is insignificant in the wideband limit.
Our new result involves an upper bound on the SNR, below which the channel uncertainty penalty prevents communications.

\newcommand{\sq}[1]{\left\| {#1} \right\|^2}
\newcommand{\itok}{^{i\rightarrow k}}
\newcommand{\sSNR}{\sqrt{\SNR}\,}
\newcommand{\Ht}{\tilde{H}}
\section{Model}
\label{sec:model}
We consider communication systems with a (single sided) bandwidth $W$, operating over block-coherent multipath channels.
We use a real discrete model of the system, after sampling at the receiver at rate $W$.
The model over a single coherence period is given by
\begin{equation}
Y=\sSNR       X \star\tilde{H}+Z \label{eq:model1}
\end{equation}
where $Y$ is the received signal over an entire coherence period of length $T_c$, this is a vector with $K_c=T_cW$ entries.
The vector $X$ of length $K_c$ and average energy $K_c$ represents the transmitted signal, the multipath channel is represented by the vector $\tilde{H}$ of length $K_c$ and $\star$ marks a convolution.
$Z$ is white standard Gaussian noise (IID with zero mean and variance one) and $\SNR$ is the signal to noise ratio.
The $\SNR$ can be understood as the signal to noise ratio per frequency resolution bin or per degree of freedom.
We neglect in~(\ref{eq:model1}) the edge effects at the beginning of the coherence period.

We impose a probabilistic energy constraint: 
\begin{equation}
P\left(\sq{X}>\left(1+o(1)\right) K_c\right)\xrightarrow[W\rightarrow\infty]{}0 \label{eq:condEnergy}
\end{equation}
Note that IID signaling satisfies this assumption.

The transmitted signal may be impulsive, we consider the signal to noise ratio during active transmission so there is no need to explicitly address the impulsiveness used by the system (i.e.\ the duty cycle ratio).
We assume that the transmitter does not use information on the channel realization, and the transmitted signal does not depend on it.

The transmitted signal $X$ is wideband: its empirical auto-correlation is upper bounded by
\begin{equation}
\left|\left<\overline{X^i},\overline{X^j}\right>\right|\leq B_4\sqrt{K_c}\ \ \ \ \ i\neq j\ \ \  i,\ j=1,\dots,K_c
\end{equation}
with $B_4$ is a constant that does not depend on the bandwidth.
The notation $<\ ,\ >$ is used for the inner product of vectors, and the notation $\overline{X^i}$ is used for a vector $X$ that is cyclicly shifted by $i$ positions, i.e.\
\begin{equation}
\overline{X^i}=\left(\begin{array}{c}
X_{\left(1-i\right)} \\
X_{\left(2-i\right)} \\
\vdots \\
X_{\left(K_c-i\right)} 
\end{array}\right)
\end{equation}
where $(\ -\ )$ indicates a mod~$K_c$ difference.


The channel is composed of $L$ paths, each with a delay in the range $\left[0,T_dW\right]$ where $T_d$ is the delay spread.
The channel is block-constant with coherence time $T_c$, i.e.\ it has IID realizations over different coherence periods.
We assume \mbox{$T_d\ll T_c$} and thus justify to an extent our loose treatment of edge effects at the beginning of each coherence period.
We approximate~(\ref{eq:model1}) with a circularly-shifted matrix:
\begin{equation}
Y=\sSNR      x\tilde{H}+Z \label{eq:model3}
\end{equation}
where
\begin{eqnarray}
x & = & \left(\begin{array}{cccc} 
X_1 & X_{K_c} & \dots & X_2 \\
X_2 & X_1 & X_{K_c} & \dots \\
\vdots  & & \ddots & \ddots \\
X_{K_c} & \dots & X_2 & X_1
\end{array}\right)  \nonumber \\
& = & \left(\begin{array}{cccc} 
\overline{X^{0}} &
\overline{X^{1}} &
\dots &
\overline{X^{K_c-1}} 
\end{array}\right)
 \label{eq:x}
\end{eqnarray}

The channel model is real, $L$ channel gains are IID and zero mean, with variance $1/L$, so the energy in the channel's impulse response equals one on average.
We assume an upper bound on path gains $\left|H_i\right|>\frac{B_1}{\sqrt{L}}$, with a constant $B_1$ that does not depend on the bandwidth.
The choice of the $L$ non zero taps is uniform over the $K_c\choose L$ possibilities.

The number of paths $L$ diverges as the bandwidth increases in a sub-linear manner~\cite{saadane_2004,saadane_2005}, i.e.\ $L\xrightarrow[W\rightarrow\infty]{}\infty$ and $L/W\xrightarrow[W\rightarrow\infty]{}0$.

We make a probabilistic assumption on the channel's response
\begin{equation}
P\left(\left|\sum_{j=1,\ j\neq i}^{K_c}\tilde{H}_j\left<\overline{X^i},\overline{X^j}\right>\right|> B_3\sqrt{K_c}\right)\xrightarrow[W\rightarrow\infty]{}0\ \ \ i=1,2,\dots,K_c \label{eq:cond}
\end{equation}
with a constant $B_3$ that does not depend on the bandwidth.
The typical value of the correlation in~(\ref{eq:cond}) is $\sqrt{K_c}$, so this assupmtion is a natural one.
By taking a large constant $B_3$ we ensure that our result holds for almost all values of $i$.

\section{Result}
\label{sec:result}
\begin{Theorem}
\label{thm}
Communication systems modeled by~(\ref{eq:model3}), that use spreading signals and operate over multipath channels as described in Section~\ref{sec:model}, with $\SNR\ll \frac{\log{W/L}}{W/L}$ have a diminishing rate in the limit of large bandwidth:
\begin{equation}
\frac{I(Y;x)}{\frac{1}{2}K_c\SNR}\xrightarrow[W\rightarrow\infty]{}0 \nonumber
\end{equation}
with probability 1.
\end{Theorem}
We prove the theorem in Section~\ref{sec:proof} by showing that 
\begin{equation}
\lim_{W\rightarrow\infty}\frac{I\left(Y;\tilde{H}|x\right)}{\frac{1}{2}K_c\SNR}=1 \nonumber
\end{equation}
and applying
\begin{eqnarray}
I\left(Y;x\right) & = & I\left(Y;\tilde{H}, x\right)-I\left(Y;\tilde{H}|x\right) \label{eq:I2} \\
& \leq & \frac{1}{2}K_c\SNR-I\left(Y;\tilde{H}|x\right) \label{eq:Iub}
\end{eqnarray}
The term $I\left(Y;\tilde{H}|x\right)$ in~(\ref{eq:Iub}) is the datarate penalty due to channel uncertainty.

\section{Discussion}
\label{sec:discussion}
The proof of Theorem~\ref{thm} is based on calculating the mmse estimate of the channel response $\tilde{H}$, given the transmitted and the received signals.
We show that this mmse estimate is a vector with an $o(1)$ norm in low SNR conditions, essentially because the noise $Z$ overwhelms the information carrying signal.

Theorem~\ref{thm} shows that in the wideband limit, the low SNR regime can be divided in to parts: 
\begin{equation}
\SNR\ll \frac{\log{K_c/L}}{K_c/L}
\end{equation}
where spreading signals are not effective, and 
\begin{equation}
\SNR> \frac{\log{K_c/L}}{K_c/L}
\end{equation}
where although the SNR diminishes in the limit, it enables a positive datarate.

\begin{figure}
\begin{center}
\leavevmode
\epsfxsize=\figs\epsfbox{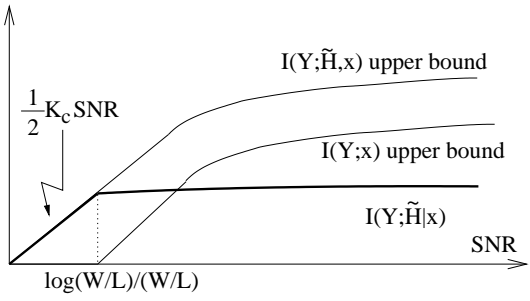}
\caption{
A sketch of the I-SNR relationship for spread signals for a very large bandwidth.
The coherent datarate upper bound (top graph) is linear in the low SNR regime, and convex.
The channel uncertainty penalty in the bottom graph is linear for low SNR values, and saturates at the channel's entropy.
The incoherent datarate (middle graph) is not convex.
}
\label{fig:ISNR}
\end{center}
\end{figure}

The channel uncertainty penalty is upper bounded by the channel entropy, and the bottom graph of~Figure~\ref{fig:ISNR} thus saturates at $\SNR=\frac{\log{K_c/L}}{K_c/L}+\frac{o\left(L\right)}{K_c}$, where the first part corresponds to he entropy of the paths' delays and the second to their gains.

\section{Proof of Theorem~\ref{thm}}
\label{sec:proof}
The proof is based on the I-mmse connection, in particular Theorem~2 of~\cite{guo_2005}.
This theorem gives a simple formula to the achievable rate of communications over a known vector channel in terms of the error of the mmse estimate of the transmitted signal.
We reverse the roles of $H$ and $x$ in our usage of Theorem~2 of~\cite{guo_2005}, i.e.\ consider $x$ as known and $H$ as the estimated party.
Using our notation, the I-mmse connection says that as long as the vector $H$ satisfies E$\sq{H}<\infty$ we have
\begin{equation}
I\left(H;\sSNR      xH+Z\right)=\frac{1}{2}\int_0^\SNR \mmse\left(\SNR\right) d\SNR \label{eq:I}
\end{equation}
where $\mmse\left(\SNR\right)$ is given by
\begin{equation}
\mmse\left(\SNR\right)=E\left[\sq{x\tilde{H}-x\hat{H}\left(Y;\SNR\right)}\right] \label{eq:mmse}
\end{equation}
and $\hat{H}$ is the mmse estimate of $\tilde{H}$ given both $x$ and $Y$.
We will show that the mmse estimate is a vector with an $o(1)$ norm in low SNR conditions, thus the minimal mean square error~(\ref{eq:mmse}) converges to $K_c$, the mutual information~(\ref{eq:I}) converges to $\frac{1}{2}K_c\SNR$ and~(\ref{eq:Iub}) diminishes.

The mmse estimate is given by
\begin{equation}
\hat{H}=E\left[H|Y,x\right] \label{eq:Hhat0}
\end{equation}
We lower bound the mmse by calculating the minimum mean square error in a system that is given additional information on $\tilde{H}$, namely which of its positions satisfies~(\ref{eq:cond}).
\begin{equation}
\hat{H}'=E\left[H|Y,x,I_{(\ref{eq:cond})}\right] \label{eq:Hhat1}
\end{equation}
where~$I_{(\ref{eq:cond})}$ is a list of indices $\left\{i\right\}$ where $\Ht_i$ satisfies~(\ref{eq:cond}). 
The additional information can only reduce the mmse.

\begin{equation}
\hat{H}'=\int\int\dots\int Hf\left(H|Y,x\right)dH_1\ dH_2\dots dH_L \label{eq:Hhat2}
\end{equation} 

The conditional probability density in~(\ref{eq:Hhat2}) is manipulated  using the independence of the transmitted signal from the channel.
\begin{equation} 
\hat{H}'  =  
\frac{\int\int\dots\int Hf\left(Y|x,H\right)f\left(H\right)dH_1\ dH_2\dots dH_L}
{\int\int\dots\int f\left(Y|x,H\right)f\left(H\right)dH_1\ dH_2\dots dH_L} \label{eq:Hhat4}
\end{equation} 
The conditional probability density in~(\ref{eq:Hhat4}) is Gaussian as $Y|x,H\sim \mathcal{N}\left(xH,I\right)$.
We denote by $f_s(\ )$ the probability density of a $K_c$ long vector of IID standard Gaussian variables
\begin{equation}
f_s(S)=\frac{1}{\left(2\pi\right)^{K_c/2}}{\exp\left(-\frac{1}{2}\sq{ S }\right)} 
\end{equation}

and proceed to examine the components of the vector $\hat{H}'$.
Consider first positions (indices) $j$ where $\tilde{H}_j=0$.
At these positions, any non-zero value of $\hat{H}_j'$ increases the estimation error and can be disregarded in the calculation of a lower bound on the mmse.
Let us examine the positions where $\tilde{H}_i\neq 0$ and~(\ref{eq:cond}) holds, and look at the estimates of each such value:
\begin{eqnarray} 
\hat{H}_i'   & =   
\int\int\dots\int & H_if_s\left(\sSNR      x\left(\tilde{H}-H\right)+Z\right) \nonumber \\
&& f\left(H\right)dH_1\ dH_2\dots dH_L\ {\LARGE /} \nonumber \\
& \int\int\dots\int & f_s\left(\sSNR      x\left(\tilde{H}-H\right)+Z\right) \nonumber \\
&& f\left(H\right)dH_1\ dH_2\dots dH_L  \label{eq:Hihat1}
\end{eqnarray} 
We prove that the mmse estimate~(\ref{eq:Hihat1}) is very small
by comparing the integral in the nominator to the integral in the denominator, that effectively sums over a bigger group of assignments of $H$.
Assuming~(\ref{eq:cond}) we show that the nominator of~(\ref{eq:Hihat1}) is negligible when compared to its denominator.

We first approximate both integrals in~(\ref{eq:Hihat1}) by sums over sampled groups of values of each positions in the vectors $H$.
The sampling is done over a tight enough grid that the resulting errors are small.

An upper bound to~(\ref{eq:Hihat1}) is calculated by breaking the sum in the denominator to a series of sums over groups of values of $H$, where each group corresponds to a single assignment of $H$ in the nominator.
The set $\mathcal{B}$ consists of assignments with a non-zero value $H_i\neq 0$.
Rewriting the discrete approximation of~(\ref{eq:Hihat1}) we get
\begin{equation} 
\hat{H}_i'  =  
\frac{\sum_{H\in \mathcal{B}} H_if_s\left(\sSNR      x\left(\tilde{H}-H\right)+Z\right)p\left(H\right)}
{\sum_{G \in \mathcal{B}+\mathcal{B}^c} f_s\left(\sSNR      x\left(\tilde{H}-G\right)+Z\right)p\left(G\right)}  \label{eq:Hihat2}
\end{equation} 
the notation $G$ was introduced to improve clarity we define $p(H)=f(H)\Delta H^L$ and $\Delta H$ is the sampling step of each dimension of the vector $H$.

We proceed to divide the entire range of vectors $G$ into non-overlapping subgroups, such that for each $H\in \mathcal{B}$ we have a corresponding subgroup $\mathcal{A}\left(H\right)$, such that 
\begin{equation} 
J(H)=\frac{H_if_s\left(x\left(\tilde{H}-H\right)+Z\right)p\left(H\right)}
{\sum_{G\in \mathcal{A}\left(H\right)} f_s\left(x\left(\tilde{H}-G\right)+Z\right)p\left(G\right)}  \label{eq:JH}
\end{equation} 
diminishes in the limit of large bandwidth.
The convergence of~(\ref{eq:Hihat2}) to zero follows directly, as the nominator of~(\ref{eq:Hihat2}) is a sum of the nominators of $J(H)$ for all $H\in\mathcal{B}$.

The subgroups $\mathcal{A}(H)$ are created randomly.
For each assignment of $G$ that has a non-zero value in the \thd{$i$} position, it is put in the subgroup $\mathcal{A}(G)$.
For a vector $G$ with $G_i=0$ we (uniformly) choose one of its non-zero taps and replace it to the \thd{$i$} position.
To clarify the process, let us say that the \thd{$j$} position of the vector $G$ was chosen.
We calculate a new vector $H$ by
\begin{equation}
H_i=G_j\ ; \ \ 
H_j=0 \ ; \ \ 
H_k=G_k \ \text{for}\ k\neq i,\ j
\end{equation}
and assign the vector $G$ to the subgroup $\mathcal{A}(H)$.

Each group $\mathcal{A}(H)$ contains $H$ and about $\left(K_c-L\right)/L$ other members, each different from $H$ in exactly two positions.
We ensure that groups' sizes do not deviate significantly from $\left(K_c-L\right)/L$ by relocating memebers from large groups into suitable smaller ones.

We denote by $H^{i\rightarrow k}$ a member of $\mathcal{A}(H)$ that differs from $H$ by exchanging the values in its \thd{$i$} and \thd{$k$} positions, and define $H^{i\rightarrow i}=H$.
The set $\mathcal{K}(H)$ holds the values of $k$ such that $H\itok\in\mathcal{A}(H)$.

The terms $p\left(H\right)$ in the nominator of~(\ref{eq:JH}) and $p\left(G\right)$ in the denominator are identical for all members of the group $\mathcal{A}(H)$ because of our assumptions on IID gains and a uniform spread of the path delays.
\begin{equation} 
J(H) =  \frac{H_i\exp\left\{-\frac{1}{2}\sq{Y-\sSNR      xH^{i\rightarrow i}}\right\}}
{\sum_{k\in \mathcal{K}(H)}\exp\left\{-\frac{1}{2}\sq{Y-\sSNR      xH\itok}\right\}} \label{eq:J2}
\end{equation} 
The denominator of~(\ref{eq:J2}) contains a sum over about $K_c/L$ exponents with different values of $k$, including $k=i$ and the nominator holds a signal such factor with $k=i$.
We take a close look at their exponent 
and introduce the notation $I\left(H_i,k\right)$ for a $K_c$-long vector with the value $H_i$ at the \thd{$k$} positions and zeros elsewhere.
\begin{eqnarray}
& - & \frac{1}{2}\sq{Y-\sSNR      xH^{i\rightarrow k}}   =  \nonumber \\
& - & \frac{1}{2}\sq{Y} \label{eq:Jl1} \\
& - & \frac{1}{2}\sq{\sSNR      x\left(H\itok-I\left(H_i,k\right)\right)} \label{eq:Jl2} \\
& - & \frac{1}{2}\sq{\sSNR      xI\left(H_i,k\right)}\label{eq:Jl3} \\
& - & \left<
{\sSNR      x\left(H\itok-I\left(H_i,k\right)\right)},\right. \nonumber\\
&& \left.{\sSNR      xI\left(H_i,k\right)}
\right> \label{eq:Jl4} \\
& + & \left<\sSNR      x\tilde{H},\sSNR      x\left(H\itok-I\left(H_i,k\right)\right)\right> \label{eq:Jl5} \\
& + & \left<\sSNR      x\tilde{H},\sSNR      xI\left(H_i,k\right)\right> \label{eq:Jl6} \\
& + & \left<Z,\sSNR      x\left(H\itok-I\left(H_i,k\right)\right)\right> \label{eq:Jl7} \\
& + & \left<Z,\sSNR      xI\left(H_i,k\right)\right> \label{eq:Jl8}
\end{eqnarray}
we now deal with each line~(\ref{eq:Jl1})-(\ref{eq:Jl8}) separately, to show that the nominator of~(\ref{eq:J2}) is much smaller than the denominator.

\paragraph*{(\ref{eq:Jl1})} The term $-\frac{1}{2}\sq{Y}$ does not depend on $k$.
\paragraph*{(\ref{eq:Jl2}), (\ref{eq:Jl5}), (\ref{eq:Jl7})}
These terms do not depend on $k$ because the vectors $H\itok-I\left(H_i,k\right)$ are identical over \mbox{$k\in \mathcal{K}(H)$}.
\paragraph*{(\ref{eq:Jl3})} The term $-\frac{1}{2}\sq{\sSNR      xI\left(H_i,k\right)}$ depends on $k$, but the norm is constant over $k\in\mathcal{K}(H)$.
The vector $I\left(H_i,k\right)$ essentially extracts a single column of the matrix $x$ and multiplies it by $H_i$.
The matrix $x$ is circularly symmetric and thus~(\ref{eq:Jl3}) is fixed.
\paragraph*{(\ref{eq:Jl4})} The term 
\begin{eqnarray}
a_k& = & -\left<
{\sSNR      x\left(H\itok-I\left(H_i,k\right)\right)},\right.  \nonumber \\
&& \left. {\sSNR      xI\left(H_i,k\right)}
\right> \nonumber \\
& = & -\SNR H_i\sum_{j=1\ j\neq k}^{K_c}H_j\left<\overline{X^{j-1}},\overline{X^{k-1}}\right> \nonumber
\end{eqnarray} 
is significantly smaller than~(\ref{eq:Jl6}) at the nominator of~(\ref{eq:J2}), or in other words an order of magnitude smaller than $\frac{K_c\SNR}{L}$.
We prove this by calculating
\begin{equation}
a_i=-\SNR H_i \sum_{j=1,\\ j\neq i}^{K_c} H_j \left<\overline{X^{j-1}},\overline{X^{i-1}}\right> \label{eq:Hxcorr}
\end{equation}
The typical value is on the order of $B_1\SNR\frac{\sqrt{K_c}}{\sqrt{L}}$ and a loose upper bound is given by 
Using condition~(\ref{eq:cond}) on $\tilde{H}$, that basically ensures a low correlation in~(\ref{eq:Hxcorr}), we have
\begin{equation}
\label{eq:a_ibound}
\left|a_i\right|\leq B_1B_3\SNR\sqrt{\frac{K_c}{L}}
\end{equation}
 
\paragraph*{(\ref{eq:Jl6})} The term 
\begin{eqnarray}
b_k & = & \left<\sSNR      x\tilde{H},\sSNR      xI\left(H_i,k\right)\right> \nonumber \\
& = & \SNR H_i\sum_{j=1}^{K_c} \tilde{H}_j\left<\overline{X^{j-1}},\overline{X^{k-1}}\right> \nonumber
\end{eqnarray}
is the dominant term in the nominator of~(\ref{eq:J2}), i.e.\ for $k=i$.
\begin{equation}
b_i=\SNR H_i \sum_{j=1}^{K_c} \tilde{H}_j \left<\overline{X^{j-1}},\overline{X^{i-1}}\right>
\end{equation}
The dominant term in the sum is the \thd{i}, where $\left<\overline{X^{i-1}},\overline{X^{i-1}}\right>\leq \left(1+o(1)\right) K_c$ with high probability.
\begin{eqnarray}
\left|b_i\right| & \leq & B_1^2 \left(1+o(1)\right) \SNR \frac{K_c}{L} +B_1B_3\SNR\frac{\sqrt{K_c}}{\sqrt{L}}  \nonumber \\
& = &B_1^2\left(1+o(1)\right)\SNR\frac{K_c}{L}
\end{eqnarray}
the last approximate equality is tight in the limit of large bandwidth.
\paragraph*{(\ref{eq:Jl8})} The term $c_k=\left<Z,\sSNR      xI\left(H_i,k\right)\right>$ is the dominant term in the denominator of~(\ref{eq:J2}).
The sum of exponents of $\left\{c_k\right\}$ is lower bounded by a single exponent with $k^\star\in\mathcal{K}(H)$.
We use asymptotic order statistics to show that there is ${k^\star}\in\mathcal{K}(H)$ such that $c_{k^\star}=\sqrt{\frac{K_c\SNR}{L}}\sqrt{2\log\frac{K_c}{L}}$ in the limit.
To prove the existance of $k^{\star}$ we examine the joint probability density of $\left\{c_k\right\}$.
These are mutually Gaussian zero mean random variables, with variance
\begin{equation} \var\left(c_k\right) =H_i^2 \SNR\left<\overline{X^k},\overline{X^k}\right>\leq H_i^2\SNR\left(1+o(1)\right) K_c \nonumber
\end{equation}
and covariance 
\begin{equation}
\cov\left(c_k,c_m\right)= H_i^2 \SNR \left<\overline{X^k},\overline{X^m}\right>\leq H_i^2\SNR B_4\sqrt{K_c} \nonumber
\end{equation}
We collect $\left\{c_k\right\}$ into the vector $C$ of length $M=\left|\mathcal{K}(H)\right|$ and mark its corrlation matrix by $R_c$.
$R_c$ is positive definite, it has a constant and large value on its diagonal, and significantly smaller values off-diagonal.

The mean and variance of the maximal of $M$ IID $\sim N\left(0,\sigma^2\right)$ random variables are given by~\cite{cramer_book}: the mean equals 
\begin{equation}
\sigma\left(\sqrt{2\ln M}-\frac{\ln\ln M+\ln 2\pi-2\mathcal{C}}{2\sqrt{2\ln M}}+O\left(\frac{1}{\ln M}\right)\right) \label{eq:OSmean}
\end{equation}
and the variance is
\begin{equation}
\frac{\pi^2\sigma^2}{12\ln M}+O\left(\frac{1}{\ln ^2M}\right) \label{eq:OSvar}
\end{equation}
where $\mathcal{C}\approx0.5772$ is Euler's constant.
Note that for a large $M$ the variance diminishes.
These results cannot be directly applied to the maximal $\left\{c_k\right\}$ because these variables are correleated.
We show that the correlations among $\left\{c_k\right\}$ are insignificant in the limit of large bandwidth in the sense that there is a $c_{k^\star}$ that is very similar to the maximal of IID Guassians,
and coclude that in the limit
\begin{eqnarray}
c_{k^\star} & \rightarrow &
(1+o(1))\sqrt{H_i^2\SNR K_c} \sqrt{2\ln M}
\end{eqnarray} 

In the nominator of~(\ref{eq:J2}) the term $c_i$ is insignificant, it has zero mean and a small variance on the order of $\sqrt\frac{K_c\SNR}{L}$.

To summarize the discussion of~(\ref{eq:Jl1})-(\ref{eq:Jl8}), we can upper bound~(\ref{eq:J2}) in the limit of large bandwidth using the significant terms in the nominator and denominator:
\begin{eqnarray} 
J(H) & \leq  & 
\frac{H_i\exp\left\{-\frac{1}{2}\sq{Y-\sSNR xH^{i\rightarrow i}}\right\}}
{\max_{k\in\mathcal{K}(H)}\exp\left\{-\frac{1}{2}\sq{Y-\sSNR xH\itok}\right\}} \nonumber\\
& \approx & H_i\exp\left\{a_i+b_i-c_{k^\star}\right\} \\
& \leq &
 H_i\exp\left\{3 B_1^2\frac{K_c\SNR}{L}-
\sqrt{\frac{K_c\SNR}{L}}\sqrt{2\log\frac{K_c}{L}}
\right\} \nonumber \\
&&
 \label{eq:exp}
\end{eqnarray} 
and for $\SNR=o\left(\frac{\log\frac{K_c}{L}}{\frac{K_c}{L}}\right)$ the exponent of~(\ref{eq:exp}) diverges to $-\infty$ as the bandwidth increases, and $J(H)\xrightarrow[W\rightarrow\infty]{}0$.
Replacing $K_c$ by $WT_C$, the proof of Theorem~\ref{thm} is complete.

\bibliography{refs}
\end{document}